# Noise and Thermal Performance of a Sub-Attofarad Capacitance Sensor for Precision Measurements, with Applications in Gravitational Wave Detectors

S. Saraf[1], S. Buchman[2], C.Y. Lui[1], S. Wang[1,3], J. Lipa[2]

[1]*SN&N Electronics, Inc., 1846 Stone Avenue, San Jose, CA 95125 USA*
[2]*Hansen Experimental Physics Laboratory, Stanford, CA 95038 USA*
[3]*Hainan Tropical Ocean University, Sanya, 572022 China*

**Abstract**

We describe the design principles, fabrication, and characterization of a precision AC resonant capacitance bridge (RCB) sensor, based on a resonant differential planar printed circuit board transformer with a solid (ungapped) MnZn ferrite core, demonstrating a short-term sensitivity at 293 K of $0.225 \pm 0.005$ aF/√Hz, at around 120 kHz resonance frequency and 1 Hz Fourier measurement frequency. At 120 K the RCB short term noise sensitivity is $0.118 \pm 0.005$ aF/√Hz. We compare the ungapped configuration to five different RCBs; three with a core gap of 65 μm and two with a core gap of 130 μm. Their average room temperature short term noise sensitivities are $0.30 \pm 0.01$ aF/√Hz and $0.45 \pm 0.01$ aF/√Hz, while the cryogenic operation of these transformers at 120 K resulted in averaged sensitivities of $0.23 \pm 0.01$ aF/√Hz and of $0.36 \pm 0.01$ aF/√Hz respectively. Multi-hour room temperature runs, with one core of each of the three gap types, proved the stability of their long-term sensitivities of $0.234 \pm 0.005$ aF/√Hz, $0.338 \pm 0.009$ aF/√Hz, and $0.435 \pm 0.010$ aF/√Hz for the ungapped (40-hour duration) and the 65 μm and 130 μm (28-hour duration) cores respectively. At 0.1 mHz, a critical frequency for space gravitational wave detectors, the respective sensitivities are $0.25 \pm 0.02$ aF/√Hz, $0.35 \pm 0.02$ aF/√Hz, and $0.53 \pm 0.07$ aF/√Hz. Measurements with the ungapped transformer configuration for temperatures from 325 K to 349 K further validate the dependence of the noise model on the temperature and permeability. The performance of our RCB with an ungapped core matches the calculated performance value and shows an improvement in signal-to-noise of two or more compared with capacitance bridges developed for similar applications. A further factor of about two noise reduction is achieved by cooling to 120 K.

## I. INTRODUCTION

There are several applications that require precise and stable measurements of capacitances with resolution at the attofarad level. Accurate proximity sensing relies on capacitance measurements for applications like microscope focusing, lens alignment or stress analysis[1]. Scanning capacitance microscopy employs capacitance measurements to raster scan a probe to get an image of surface topography or local changes of dielectric constant[2]. Position sensors that infer a relative sub-nm displacement via a capacitance measurement have important applications in many space-based relativity experiments like Gravity Probe B[3] and tests of the equivalence principle[4,5]. Space-based gravitational wave antennas like the future LISA[6], Taiji[7], and TianQin[8] missions employ capacitance-based position sensors in conjunction with high voltage actuation systems for drag-free operation of a gravitational reference sensor as part of their disturbance reduction systems[9].

As is common practice, the instrument described here is based on an AC capacitance bridge with the advantage of performing high signal-to-noise measurements that are not at DC. Since the drive frequency of the bridge can be chosen as desired, resonant operation with stable inductors or transformers can be used to improve the signal-to-noise of the measurement. There are several advantages to using a resonant differential transformer. First, the signal-to-noise ratio is enhanced by the quality factor $Q$ of the resonant circuit at the operating frequency. Second, common-mode noise is nulled out to first order due to the differential operation of the transformer's primary windings. Third, due to transformer action, there is electrical isolation of the front-end electronics from the sensing electronics. This prevents formation of ground loops, while balanced differential operation of the entire circuit will reduce other forms of coupling to the precision measurement system.

To date, the state-of-the-art in capacitance sensing for a similar application was demonstrated by the LISA Pathfinder (LPF) mission[10], showing a noise performance of between 0.7 aF/√Hz and 1.8 aF/√Hz (for an excitation voltage of 0.6 V) down to a frequency of 1 mHz. Measurements showed a 2.6 spread factor between the noise figures of 12 nominally identical bridges used in LPF; two for each of the three axes of the two test masses[10]. This spread was traced to the hard-to-control variations of property details in the hardware. We emphasize that this performance was achieved as part of a complete system that included the higher voltages of the forcing function for the test masses, as well as all other noise sources contributed by the auxiliary electronics and hardware.

The hardware development, implementation, and testing of the capacitance bridge described here was performed by SN&N Electronics[11] in collaboration with Stanford University and partially funded through the Changchun Institute of Optics, Fine Mechanics, and Physics (CIOMP) of the Chinese Academy of Science and is described in detail in a number of technical reports[12,13,14]. A test by CIOMP, using the design detailed in these technical reports[15], resulted in a noise level of 1.1 aF/√Hz for frequencies of between 10 mHz and 1 Hz. Our calculations for, and results of tests with the more advanced SN&N hardware are presented in this paper, with our hardware showing a five-times better performance, 0.22 aF/√Hz, than that in reference 15.

Simulations and analyses of the practical limits for scanning capacitance microscopy[2] predict a sensitivity of 0.36 aF in a 1 kHz bandwidth measurement. However, this limit is set by constraints due to the geometry of the scanning tip rather than by the capacitance bridge design. For a top performing commercially available capacitance meter see the Andeen-Hagerling instruments[16].

Table 1 shows the performance of capacitance bridges developed for applications similar to the one described here, that is suitable for gravitational wave detectors and other low capacitance precision measurements. The first row refers to the instruments flown on the geodesy missions

GRACE[17], GRACE-FO[18], and GOCE[19] and on the Equivalence Principle measurement MICROSCOPE[20]; all developed by CNES, the French National Centre for Space Studies. Note that the capacitance to the test masses is significantly larger in these instruments, compared to the gravitational wave detectors test masses of LISA Pathfinder[21], and TianQin[8], represented in all other rows and developed by ESA, the European Space Agency, the Huazhong University of Science in China, and SN&N Electronics in collaboration with Stanford University, as described here. The capacitance resolutions given in column four are normalized to an excitation level of 1 V.

**Table 1. Performance of capacitance sensors with capabilities approaching the ones described in this paper and for the sensors developed in this work.**

| Application | Instrument Design and Manufacture | Resonant Frequency @293K (kHz) | Capacitance Resolution (aF/√Hz)×V | Comments |
|---|---|---|---|---|
| GRACE, GRACE-FO, GOCE, MICROSCOPE | CNES | 100 | 0.7[22,23] | Complete flight systems |
| LISA Pathfinder | ESA | 100 | 0.4 - 1.1[10] | Complete flight systems |
| TianQin | Huazhong U. of Science, China | 50 | 1.2[24] | Laboratory Capacitance Bridge |
| TianQin | Huazhong U. of Science, China | 200 | 0.7[25] | Laboratory Capacitance Bridge |
| Precision Meas. and GW detection **THIS WORK ungapped Core#6** | SN&N, USA | 127 | 0.22 – 0.12† | Laboratory Capacitance Bridge |
| Precision Meas. and GW detection **THIS WORK 65 μm Cores#1,4,5** | SN&N, USA | 118 | 0.33 – 0.21† | Laboratory Capacitance Bridge |
| Precision Meas. and GW detection **THIS WORK 130 μm Cores#2,3** | SN&N, USA | 118 | 0.42 – 0.34† | Laboratory Capacitance Bridge |

The paper is organized as follows. In section II we describe the design of the core of the RCB, the resonant transformer; with subsection II.A addressing the design of the planar windings, subsection II.B describing the printed circuit board containing the windings, and subsection II.C showing the loss calculation upon which the RCB design was based. The RCB circuit is described in section III and includes the measurements of the transformer in subsection III.A, of the capacitances in subsection III.B, of the short-term noise floor in subsection III.C, and of the long-term noise and its stability in subsection III.D. All calculations and results in sections II and III are for all six cores in three configurations, (one ungapped, three with 65 μm gap and two with 130 μm gap), at the 293 K. Section IV demonstrates the temperature dependent performance of the RCB, with cryogenic measurement results at 120 K and some intermediate temperatures given in section IV.A and above room temperature measurements in section IV.B. Conclusions are summarized in section V.

---

† *Measured in this work with the transformer of the SN&N RCB at 120 K. Note that the resonant frequency shifts to higher values at low temperature.*

## II. RESONANT TRANSFORMER DESIGN

Figure 1 shows a schematic of the RCB with the transformer primaries configured in differential operation mode. Ip1 and Ip2 are the primary currents, and the nominally identical transformer primary windings are configured such that the flux in the common core will cancel out when the bridge is in balance. The resonant frequency of the tuned bridge is given by:

$$f_0 = \frac{1}{2\pi\sqrt{L(T)C_{eq}}} \quad (1)$$

where $C_{eq} = C_1 + C_2 + C_{p1} + C_{p2}$ and $L(T) = L_1(T) = L_2(T) = L_3(T)$ is the inductance of the primary windings – dependent on the temperature. For the nominal case, where the capacitances to the test masses and the tuning capacitances are equal, $C_1 = C_2 = C_0$ and $C_{p1} = C_{p2} = C_p$, the equivalent capacitance for resonant operation is given by: $C_{eq} = 2(C_0 + C_p)$.

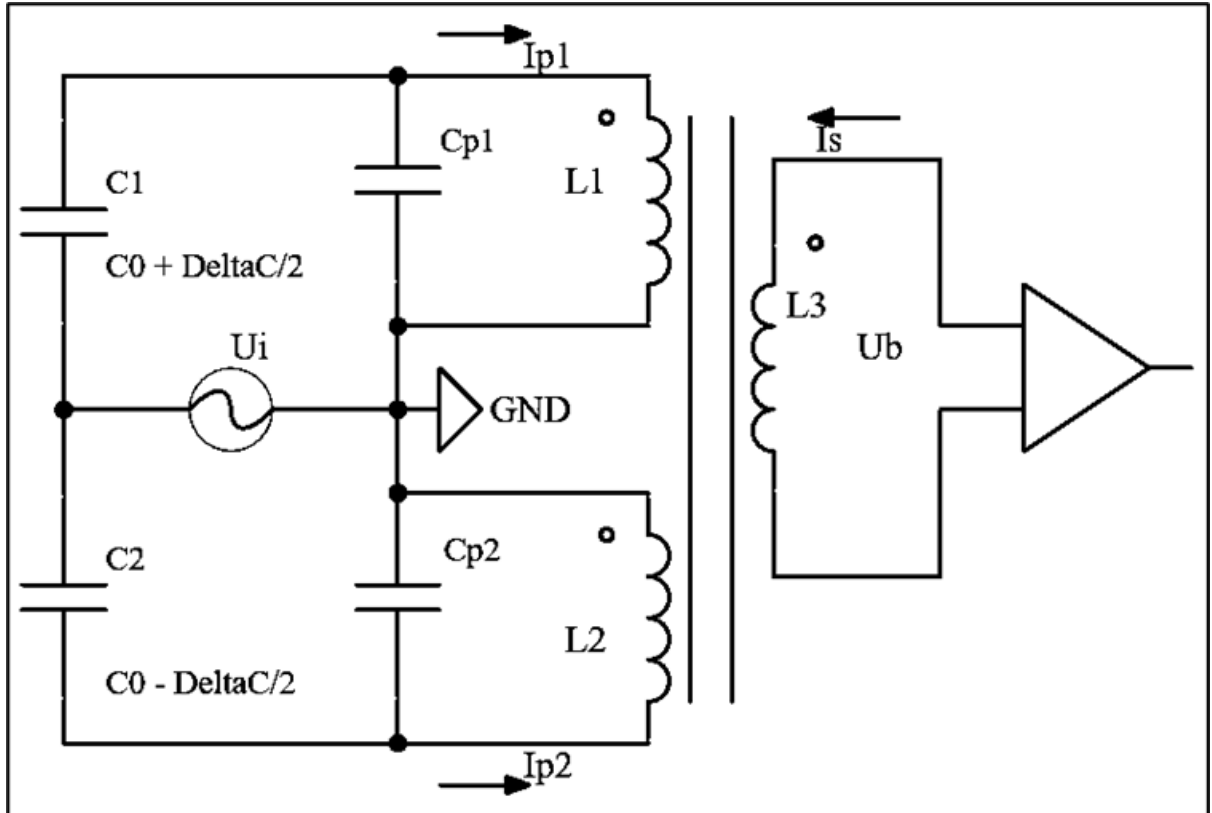

**Figure 1. Simplified diagram of a resonant capacitive bridge for noise calculations. $C_{p1}$ and $C_{p2}$ are the fixed capacitors in the bridge and serve as tuning capacitors for the resonant transformer primaries. $C_1$ and $C_2$ represent the capacitances to the test mass.**

Circuit analysis of the bridge architecture yields the following equation for the magnitude of the bridge output voltage $U_b$ at resonance frequency $f_0$:

$$|U_b(f_0)| = \frac{Q_{RCB}}{C_{eq}} \Delta C \, U_i \qquad (2)$$

where $Q_{RCB}$ is the quality factor of the RCB, $\Delta C$ is the differential capacitance between the two arms of the bridge (where for $\Delta C \ll C_{1,2}$, $\Delta C_{1,2} = \pm \Delta C/2$), and $U_i$ is the bridge actuation voltage, with typically $U_i \cong 1\text{V}$. Synchronous demodulation would follow the AC amplifier stages to extract the differential capacitance at low frequencies for further analog and digital processing. The demodulation aspects of the detector are not discussed in this paper, except to note that synchronous demodulation has a √2 noise penalty associated with it from the RF mixing process. In our analysis this factor is included to reflect the result if the demodulator was used. Standard circuit analysis shows that the equivalent capacitance noise limited by the bridge thermal noise can be written as:

$$S_C^{1/2} = \frac{1}{U_i} \times \sqrt{\frac{8k_B T}{(2\pi f_0)^3 L(T) Q_{RCB}}} \;\left(\frac{\text{F}}{\sqrt{\text{Hz}}}\right) = \frac{1}{U_i} \times \sqrt{\frac{\pi^{-3} k_B T}{FOM_{RCB}}} \;\left(\frac{\text{F}}{\sqrt{\text{Hz}}}\right) \qquad (3)$$

where $k_B$ is Boltzmann constant and $T$ is the temperature in degrees Kelvin. From equation 3 we note that the capacitance sensor noise scales inversely with excitation voltage and is proportional to the square root of the temperature and to $[L(T)Q]^{-1/2}$. Additionally, the figure-of-merit of the RCB for the measurement is defined as $FOM_{RCB} \equiv f_0^3 L(T) Q_{RCB}$. Therefore the goal is to have a high resonant frequency within the constraints of losses, bandwidth of the amplifiers, and demodulation electronics. For example, a capacitance sensing noise limit of 1 aF/√Hz at $T = 293$K for $U_i = 1$V requires that $FOM_{RCB} = 1.30 \times 10^{14}$. Operation at $f_0 \cong 2000$ kHz then requires that $L(T)Q_{RCB} \cong 0.08$ H. Later calculations will indicate that $Q_{RCB} \approx 100$ is achievable, which yields an initial required inductance of $L(293\text{K})_i \cong 0.80$ mH. We introduce a margin of safety of five for $FOM_{RCB}$, yielding $L(293\text{K}) = 4.0$ mH for the primary and the secondary inductances. Then, from equation 1, $C_{eq} = 440$ pF at the chosen 118 kHz excitation frequency, and, for the nominal $C_0 = 1$ pF, the resulting tuning capacitance is $C_p \cong 219$ pF. Consequently, we use these values as the starting point for the resonant transformer design.

An optimal design for a compact planar differential transformer would use PCB substrates made of ceramic composites[12,13,14]. The choice for the material of the core is based on a low loss and a low temperature coefficient of the complex permeability $\mu$. MnZn or NiZn ferrites are the materials of choice at operating frequencies above 100 kHz. Magnetic cores used with planar devices have a different shape than conventional cores used with helical windings. Compared to a conventional toroidal magnetic core of equal volume, devices built with optimized planar magnetic cores exhibit significantly reduced height (low profile), greater surface area for improved heat dissipation capability, larger magnetic cross-section area thus enabling fewer turns, and smaller winding area that facilitates interleaving and excellent reproducibility enabled by its winding structure. Some of the disadvantages of planar windings are the higher interwinding capacitance (that is the same for the primary and secondary windings) and the consequently lower self-resonance frequency and a lower $Q$. This can be mitigated somewhat by staggering the traces on adjacent PCB layers and by using a ceramic composite substrate with a low dielectric constant, thus resulting in a lower interwinding capacitance.

$D_\delta$, the relative loss factor for an ungapped core with initial relative (unitless) permeability $\mu_i$ and loss factor $\tan\delta_i$ is given by:

$$D_\delta = \tan\delta_i/\mu_i \tag{4}$$

All magnetic permeabilities used are relative, that is in dimensionless units.

Cores with an air gap $s$ that is small compared to the effective magnetic path length $l_e$ in the core have a lower effective permeability, $\mu_e$, that can be approximated by:

$$\mu_e = \frac{\mu_i}{1 + \frac{s}{l_e}\mu_i} \tag{5}$$

$\tan\delta_e$ is similarly given as:

$$\tan\delta_e = \tan\delta_i \frac{\mu_e}{\mu_i} \tag{6}$$

The effective temperature coefficient of the core permeability $\alpha_e$ and its quality factor are also reduced and respectively increased by the ratios of effective and initial permeabilities:

$$\alpha_e = \alpha_i \frac{\mu_e}{\mu_i}, \quad Q_{Ce} = Q_{Ci}\frac{\mu_i}{\mu_e} \tag{7}$$

The N41 MnZn ferrite core is a good candidate for our application, due to its low temperature coefficient and loss factor[26]. At room temperature the N41 MnZn principal magnetic properties are[26]: initial permeability $\mu_i = 2300 \pm 25\%$, relative loss factor $D_\delta = 4.4 \times 10^{-6}$, and resistivity $\rho = 2\,\Omega\text{m}$. The core loss factor at around 120 kHz is then: $\tan\delta_i = D_\delta\mu_i \cong 0.01$. Commonly available core shapes are Rectangular Modulus (RM) and Pot shaped (P), with the RM shape better suited to our application.

We next explore the possibility of using cores with different gaps to decrease $\mu_e$ and therefore reduce the temperature coefficient and increase the quality factor; see equation 7. The N41 MnZn ferrite core model B65813A[27], with air gap $s = 130\,\mu\text{m}$, has an effective magnetic path length $l_e = 44$ mm. Note that each of the B65813A core halves has a 65 μm air gap, thus allowing for assemblies of 130 μm gap cores, from two components of B65813A or of 65 μm gap cores by combining a B65813A core half with a matching ungapped component; see Figure 2.

Finally, the inductance L is given by:

$$L = A_L N^2 \tag{8}$$

where N is the number of windings and $A_L = 630$ nH is the inductance factor for the RM10 shaped core with the full gap of 130 μm. As the desired inductance for the bridge primary is $\cong 4.0$ mH, we choose $N = 80$, to allow for an integer number of turns per layer and resulting in a winding inductance $L = 4.03$ mH. The winding Equivalent Series Resistance (ESR), $R_s$, and the parallel resistance from core losses, $R_{P1}$, are then calculated from:

$$R_s = \frac{2\pi f_0 L}{Q_C}, \quad R_{P1} = R_s Q_C^2 \tag{9}$$

We next still need to consider the $I^2R$ losses in the copper windings/traces from AC resistance (skin and proximity effects) and loss enhancement from the self-resonant frequency, which will reduce the quality factor of the transformer. The initial effective temperature coefficient of the solid ungapped core, $\alpha_i = 800$ ppb/K, will also be reduced for the gapped and half gapped configurations by the ratio $\mu_{eff}/\mu_i$ (equation 7).

In a differential transformer, where two similar primary windings surround a common core, the temperature coefficient of the core permeability to first order is nulled out. However, a zero initial DC imbalance of the inductance between the two primary windings, $(\delta L/L)_{DC}$ can fluctuate with temperature and masquerade as a change in capacitance[28]. All above properties for the RCB transformer are summarized in Table 2 for the three configurations with gaps of 0 μm, 65 μm, and 130 μm. The values for $\mu_e$, $A_L$, and $D_\delta$ are compiled from the TDK data sheets[26,27], while all others are calculated from the equations. Given that $A_L(\text{nH}) = 2.8\mu_e$ at room temperature and 118 kHz, the relative loss factor of the core, $D_\delta$, is in theory independent of core gap (equation 6), as is $R_{P1}$ (equations 7, 8, 9). Measured $Q$ values for the entire transformer, Table 4, show significantly lower values than the core ones in Table 2.

**Table 2. Calculated and manufacturer's (shaded) values at 118 kHz for the TDK N41 MnZn ferrite core and for the planar transformer with 80 Cu windings, for core gaps of 130 μm, 65 μm, and 0 μm.**

| GAP | | CORE | | | | | WINDINGS | | | |
|---|---|---|---|---|---|---|---|---|---|---|
| Type | Size (μm) | $\tan\delta_e$ | $\mu_e$ | $D_\delta$ | $Q_C$ | $\alpha_e$ (ppb/K) | $A_L$ (nH) | $L$ (mH) | $R_S$ (Ω) | $R_{P1}$ (MΩ) |
| full gap | 130 | 0.0010 | 225 | 4.4×10⁻⁶ | 1010 | 78 | 630 | 4.03 | 3.0 | 3.0 |
| half gap | 65 | 0.0018 | 400 | 4.4×10⁻⁶ | 568 | 139 | 1120 | 7.17 | 9.4 | 3.0 |
| ungapped | 0 | 0.0100 | 2300 | 4.4×10⁻⁶ | 100 | 800 | 6440 | 41.2 | 305 | 3.0 |

## II.A. PLANAR WINDING DESIGN

B65813A cores have an inner and outer diameter available for the for the planar windings of $\Phi_{Ci} = 10.9(-0.4)$mm and $\Phi_{Co} = 21.2(+0.9)$mm, Figure 2, where the maximum mechanical tolerances specified by the manufacturer are in brackets. Therefore, for sizing purposes, the PCB should have a hole of $(\Phi_i)_{max} = 10.9$mm and an outer diameter of $(\Phi_o)_{min} = 21.2$mm. Thus, the width of the circular part of the PCB should be $\{[(\Phi_o)_{min}] - [(\Phi_i)_{max}]\}/2 = 5.15$mm, with the windings located on this PCB section.

It is best to use a substrate material with a relatively low dielectric constant to minimize the interwinding capacitance, with the properties of a few suitable board materials listed in Table 3.

**Table 3. Printed circuit board properties**

| Parameter | FR4 | ISOLA Tachyon 100G | Taconic TSM-29 (PTFE) | Taconic TLX-8 (PTFE) |
|---|---|---|---|---|
| Dielectric constant $\varepsilon_r$ | 4.40 | 3.02 | 2.94 | 2.55 |
| Dissipation factor $D$ | 0.0170 | 0.0021 | 0.0015 | 0.0019 |
| CTE-X | 14ppm/K | 15 ppm/K | 23ppm/K | 9ppm/K |
| CTE-Y | 12ppm/K | 15 ppm/K | 28ppm/K | 12ppm/K |
| CTE-Z | 70ppm/K | 45 ppm/K | 78ppm/K | 140ppm/K |
| Dielectric strength | 20MV/m | 60MV/m | 42MV/m | 45MV/m |

The TLX series from Taconic[29] has the lowest $\varepsilon_r$ and dissipation factor. While the TLY[30] series has a lower dielectric constant, it is harder to procure, laminate, and temperature cycle for a multi-layer board design and therefore not shown in Table 3. TLX-8 would be our first choice for the substrate material except it has high z-axis CTE. Conversations with vendors indicate that, considering all requirements, the ISOLA Tachyon 100G material[31], with a low z-axis CTE and an acceptable $\varepsilon_r$, would be best suited to a multi-layer fabrication. However, for the current work, FR4 was used due to its wide commercial availability, as will be discussed in section III.A.

## II.B. TRANSFORMER PRINTED CIRCUIT BOARD DESIGN

We divided the 80 turns around the ferrite core into 5 turns per layer on a 16-layer PCB. The core has a winding depth (maximum height of all windings) of 12.4 mm. We partition the winding depth into 3 planar windings (2 primary and 1 secondary) each 3.1mm thick, leaving us 3.1mm for spacers between windings to reduce the primary-to-secondary capacitance value. Two ring spacers made from the same FR4 material, 1.5 mm thickness, were used as the winding separators, Figure 2. The three windings, two PCB spacers and the ferrite core were potted together using the low-outgassing adhesive MasterBond EP37-3FLFAO. The stacking order of the windings is primary 1, secondary, primary 2. Interwinding capacitance is estimated at 15 pF. The fringing effect contribution to the capacitance is negligible for the 1.5 mm spacers. A PCB that can fit around the core was selected, with a handle-like extension to enable connections to the windings, as shown in Figure 2. Each layer has 5 turns 0.9 mm wide with a 0.1 mm spacing between them. The turns are slightly staggered by about 0.1mm between the top and bottom layers of each laminate to reduce the interwinding capacitance. Blind vias across each of the laminates connect the upper set of 5 turns to the inner set on the lower layer of each laminate. Each laminate is connected to its neighboring laminate using a through-hole via located in the flat extender part of the PCB, as shown in Figure 2, to save space for the turns in the circular portion of the PCB that encircles the core. The three windings, two primary and one secondary, are identical. Each one has 8 laminates with 5 windings each, with 7 inter-laminate connections and two terminal connections (on the flat extender part of the PCB), that produce a 4.03 mH inductor; see Figure 2.

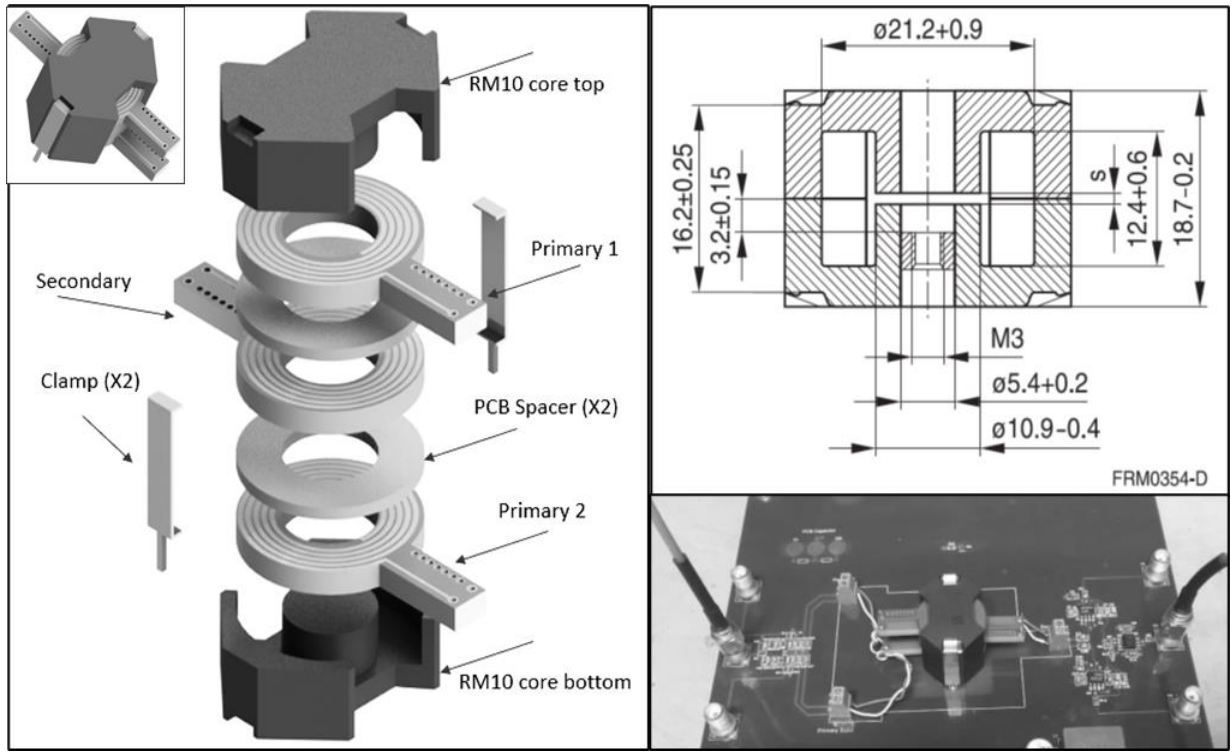

**Figure 2.** *Left*: Exploded CAD view of the ferrite core, transformer windings, spacers, and clamps. The secondary winding is sandwiched between the two primary windings. The stack-up consists of identical 3.1 mm-thick primary and secondary winding PCBs separated by 1.5 mm spacers that are made from the same PCB material, and potted and clamped around the N41 ferrite core, *Insert*: CAD perspective of assembled transformer. *Right top*: TDK dimensions for the B65813A N41 MnZn ferrite core. The cores in the presented work do not have a central hole. *Right bottom*: Photograph of transformer on RCB board.

Exchanging the windings in the stack-up before potting the assembly changed the measured inductances by $< 0.02$ mH or $< 0.5\%$, thus demonstrating that good reproducibility is achieved between all the windings using the printed circuit board technology and making apparent the advantage of the planar design.

## II.C. LOSS CALCULATIONS

The length of an 80-turn winding in the geometry shown in Figure 2 is about 4032 mm. The average width of each turn is 0.9 mm. For the $h = 35$ µm thickness of the PCB traces (fabricated from 1 oz copper) and using $\rho_{Cu} = 1.68 \times 10^{-8}$ Ωm for the resistivity of copper at room temperature, we obtain: $R_{DC} = 2.15\Omega$.

The distributed interwinding capacitance estimation is based on calculating the total energy associated with the electric field in the 2 layers and extracting the equivalent capacitance. The capacitance of a laminate $C_w$ with $n$ turns on each side can be estimated from the formula[32]:

$$C_w = \frac{(n+1)(2n+1)}{6n} C_0 \quad (10)$$

where $C_0 = (\varepsilon_r \varepsilon_0 / h_l) A_T$ is the capacitance of a single overlapping turn, $h_l$ is the thickness of the core dielectric material in a single laminate, and $A_T$ is the overlapping area of a single turn.

Fringing effects are calculated[33] to be <0.2% for the 100 µm thick laminate and can be neglected. Series connection of $m$ laminates results in a distributed winding capacitance $C_d$ that is given by:

$$C_d = \frac{4(m-1)}{m^2} C_w \quad (11)$$

Therefore, for 5 turns per laminate with a 100µm laminate thickness and 8 laminates per winding, the distributed interwinding capacitance $C_d$ is estimated at ~15 pF for a dielectric material with $\varepsilon_r = 2.55$, like the Taconic TLX-8 (PTFE). The self-resonant frequency of the winding can now be calculated from:

$$f_{sr} = \frac{1}{2\pi\sqrt{LC_d}} = 650 \text{kHz} \quad (12)$$

The apparent $Q$ at the bridge drive frequency of $f_0 = 118$ kHz will drop to $\cong 0.96Q$, a 4% reduction. The equivalent additional resistance $R_{sr}$ from the coil self-resonance is:

$$R_{sr} = 2 \times R_{DC} \times \left(\frac{f_0}{f_{sr}}\right)^2 = 0.15\Omega \quad (13)$$

Dielectric losses due to the lossy substrate are estimated using the formula:

$$R_d = 2\pi f_0 L \times \tan(\delta_d) \times \left(\frac{f_0}{f_{sr}}\right)^2 = 0.12\Omega \quad (14)$$

Finally, we estimate the AC losses from the skin and proximity effects. At 118 kHz, 1 oz copper has a skin depth $\delta_S = 188$ µm ($h = 35$µm thickness). The parameter for estimating eddy and proximity effect losses is $\xi = h/\delta_S = 0.186$. The expression for AC resistance of the $m^{th}$ layer that incorporates the skin and proximity effects is given by[32,34]:

$$\frac{R_{AC,m}}{R_{DC,m}} = \frac{\xi}{2}\left[\frac{\sinh\xi + \sin\xi}{\cosh\xi - \cos\xi} + (2q-1)^2 \frac{\sinh\xi - \sin\xi}{\cosh\xi + \cos\xi}\right] = 1.08 \quad (15)$$

where $q$ is defined as:

$$q = \frac{F(h)}{F(h) - F(0)} \quad (16)$$

and $F(h)$ and $F(0)$ are the Magnetomotive Forces (MMFs) at the top and bottom of a layer. For a layer count of 16, and using the skin depth of copper calculated earlier, $R_{AC,m}/R_{DC,m} = 1.08$ at 118 kHz. Therefore, the loss due to the skin and proximity effects is modelled as an extra resistance $R_{AC} = 0.07\Omega$. Adding all the losses, we get an equivalent loss resistance:

$$R_{tot} = R_{DC} + R_{AC} + R_{sr} + R_d = 2.5\Omega \quad (17)$$

Adding a 50% error margin, we will use $R_{tot}^M = 3.75\Omega$ for the $Q$ calculations. Note that replacing the Taconic TLX-8 (PTFE) substrate with the FR4 one, increases only $R_{sr}$ and $R_d$ by a factor of 1.7, the ratio of their dielectric constants; an increase of about 8% in $R_{tot}$, negligeable compared to the value of $R_{tot}^M$ used in the estimates.

The resulting quality factor of the winding is $Q_w = 790$ and the equivalent parallel resistance $R_{P2} = 2.46$ M$\Omega$. $R_{P2}$ in parallel with $R_{P1}$, the calculated core loss resistances calculated earlier ($R_{P1} = 0.30$ M$\Omega$, 1.31 M$\Omega$, 2.33 M$\Omega$ for no gap, half gap, full gap; see Table 2), yields $R_B$, the total effective resistance of the bridge including core, copper, and dielectric losses, that equals respectively $R_B = 0.27$ M$\Omega$, 0.85 M$\Omega$, 1.20 M$\Omega$ for no gap, half gap, and full gap cores.

The net effective quality factor of the bridge, $Q_{RCB}$, with all losses factored in, is given by:

$$Q_{RCB} = \frac{2\pi f_0 L}{R_{tot}^M + R_S} \quad (18)$$

and using the $R_S$ values calculated in Table 2 we obtain: $Q_{RCB}^{0\mu} = 89$, $Q_{RCB}^{65\mu} = 283$, $Q_{RCB}^{130\mu} = 395$ for cores with no gap, half gap, and full gap.

The tuning capacitance required to assure resonant operation at the 118 kHz bridge excitation frequency is obtained from equation 1: $C_{eq} = 436$ pF. Since $C_{eq} = 2(C_0 + C_p)$, and accounting for an interwinding capacitance of $C_0 = 15$ pF, the required tuning capacitance $C_p \cong 202$ pF. This is implemented with 180 pF and 22 pF capacitors in parallel, using devices with good temperature stability (ceramic, NP0 type). The temperature coefficient of the interwinding capacitance is to first order given by the z-axis CTE of the substrate.

For the FR4 substrate, the z-axis temperature coefficient is 70 ppm/K while the temperature coefficient of a high-quality ceramic NP0 capacitor is typically 10 ppm/K. The composite temperature coefficient of the tuning capacitance can be estimated at 20 ppm/K. Under these idealized assumptions, the resulting shift in the resonant frequency would be ≤ 100 ppm/K and therefore the detuning from the 120 kHz resonance would be ≤ 12 Hz/K. The experimental result is however between 70 Hz/K and 265 Hz/K, dominated by the strong temperature dependence of the core inductance factor $A_L$; see Table 2 and references 26 and 36.

The secondary signal was found to be slightly reactive, with a phase of about half a degree with respect to the excitation signal at 118 kHz. The output amplitude and the capacitance measurement have temperature coefficients of 50 ppm and 0.20 aF/K, respectively. The thermal noise of the bridge impedance $R_B$ at resonance drives the system measurement noise to first order. Substituting the value of $L$ and $Q$ into equation 3 yields $S_C^{1/2} = 0.22$ aF/√Hz for the bridge capacitance readout noise. The amplifiers that follow the transformer add noise to the measurement, but as shown below, their contribution is expected to be much lower than the bridge thermal noise.

### III. RESONANT CAPACITANCE BRIDGE CIRCUIT

The schematic of the measurement system is shown in Figure 3. The differential transformer secondary is connected to a pair of transimpedance amplifier (TIA) stages with a feedback network $Z_{FB}$ to convert the secondary currents into proportional voltages. Circuit analysis[10] yields the transfer function at resonance as:

$$\frac{u_o}{u_i} = k \frac{2\Delta C}{C_{FB}} G_{diff} G_{ext} \tag{19}$$

where $\Delta C$ is the differential capacitance, $C_{FB} = 2.2$ pF is the feedback capacitance of the TIA stages and $k = 0.95$ is the transformer coupling factor for a 130 μm gap. $G_{diff} = 10$ is the gain of the capacitively coupled differential amplifier stage that converts the TIA outputs to a single-ended output for transfer function and noise measurements. An external amplification stage $G_{ext} = 10$ that uses a quiet LT1028 operational amplifier was connected to elevate the bridge output well above the instrument noise floor.

The readout sensitivity conversion factor is measured to be $f_{conv} = 864$ μV/fF at 1V excitation, or including the amplifiers, $f_{conv}^G = f_{conv} G_{diff} G_{ext} = 86.4$ mV/fF. The noise contributions to the TIA are calculated to be: a) 22.6 nV/√Hz from the noise gain amplification of the AD8510 JFET operational amplifiers voltage noise $e_n$, b) 2.6 nV/√Hz from the operational amplifier current noise $i_n$ flowing through the feedback impedance, and c) 34.6 nV/√Hz from the Johnson noise contribution of the real part of the feedback impedance at resonance. Summing these contributions in quadrature, $N_{TIA}$, the total TIA noise is calculated to be 41.4 nV/√Hz and verified using the LTSpice simulation code: $N_{TIA}^{LTSpice} = 41.4$ nV/√Hz. $N_{trans}$, the transformer thermal noise at the output of the TIA stage is $N_{trans} = 134$ nV/√Hz, and is clearly the dominant source of noise. Therefore $N_{RCB}^{calc}$, the calculated total noise at the output of the RCB, including the differential amplifier, the external gain stages, and factoring in the $f_{dem} = \sqrt{2}$ demodulation penalty (as discussed earlier) is:

$$N_{RCB}^{calc} = \sqrt{N_{TIA}^2 + N_{trans}^2} \times G_{diff} \times G_{ext} \times f_{dem} \cong 20 \ \mu V/\sqrt{Hz} \Rightarrow S_{min}^{1/2} \cong 0.23 \ aF/\sqrt{Hz} \tag{20}$$

Where the minimum capacitance noise floor, $S_{min}^{1/2} \cong 0.23$ aF/√Hz, is calculated for $U_i = 1V$.

### III.A. TRANSFORMER MEASUREMENTS

As mentioned above, due to availability constraints, the planar primary and secondary coils were fabricated on the FR4 substrate, rather than the preferred Taconic TLX-8 (PTFE), thus resulting in an increased dielectric constant $\varepsilon_r$ and dissipation factor $D$. The FR4 laminates-based transformer windings had good $Q$ numbers, and the final measurements were better than 0.4 aF/√Hz capacitance noise floor. The coils were stacked as shown in Figure 2 and the coil resistance, inductance and $Q$ were measured. The secondary coils had an about 10% lower measured inductance than the primary coils due to their central position in the coil stack farthest from the ferrite core base and top.

Six different cores were tested at both room and low temperatures, 293 K and 120 K. Two cores, designated as Core#2 and Core#3 were full gap, $s = 130$ μm, three cores, designated as Core#1, Core #4, and Core#5 were half gap, $s = 65$ μm, and Core#6 was ungapped, $s = 0$ μm. Core#1 was used for all measurements presented in sections III.A, III.B, and III.C. Results in sections III.D were obtained with Core#1 and Core#2, while those in section III.E were obtained with all six cores. The FR4 PCB windings (primary upper, primary lower, and secondary) were characterized using an RLC meter and the measurements are shown in Table 4, with the averages for the cores with the same gap size given in the lower part of the table. $L, Q, R_S^w, DCR, L_{leak},$ and $k$ are the inductance, quality factor, series resistance, DC resistance, leakage inductance, and windings coupling factor.

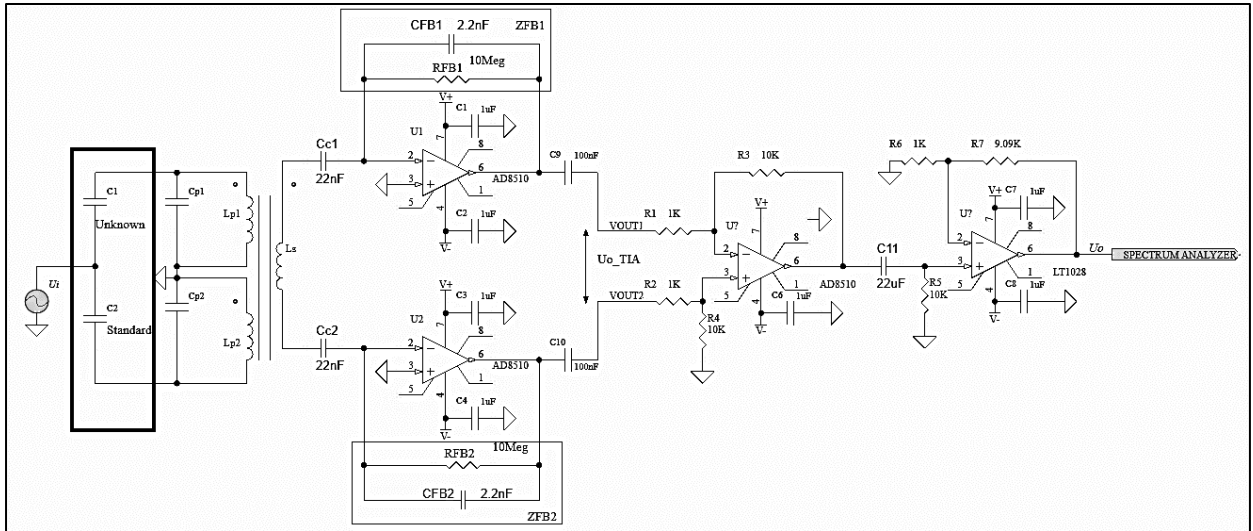

Figure 3. Schematic of the symmetrical differential transimpedance amplifier stage with a single gain setting component $C_{FB}$ per operational amplifier. $C_C$ couples the transformer to the operational amplifier and $R_{FB}$ provides a path for the operational amplifier DC bias current. $Z_{FB1}$ and $Z_{FB2}$ are the complex impedances of the feedback networks. The TIA stage is followed by a differential 10X stage and an external 10X amplifier that is connected to a spectrum analyzer for capacitance and noise floor measurements.

The leakage inductance of all three planar coils was measured by shorting the other two windings and measuring the coil inductance of the single other winding. The transformer coupling factor *k* can be calculated from these inductance measurements. All measurements were performed at the standard RLC meter measurement frequency of 100 kHz. Interchanging individual coils did not affect the measured values, indicating excellent reproducibility between the planar windings.

$C_{eq}$, the capacitance required to resonate with the measured primary inductance at the bridge excitation frequency, is calculated to be 690 pF by using equation 1. The tuning capacitances $C_{p1}$, and $C_{p2}$ required in the measurement circuit calculate to 324 pf, after allowing for the calculated winding capacitance $C_w$. These were synthesized using a 300 pf and a 24 pf C0G/NP0 ceramic capacitors in parallel. The bridge resonant frequency of 117.75 kHz was measured by driving the bridge with a function generator and noting the frequency at peak output voltage of the transimpedance stage. By adjusting the tuning capacitors, the resonance frequency can be tuned to any desired value in the range 118 ± 5 kHz without changing the RCB performance appreciably.

**III.B. CAPACITANCE MEASUREMENTS**

The capacitance bridge was excited with a 100 mV$_{rms}$ sinewave at 117.75 kHz using an HP 8648A signal generator. A differential capacitance was introduced between $C_1$ and $C_2$ and precisely measured using an Andeen-Hagerling AH 2500A[16] capacitance bridge operating at 1 kHz, and are: $C_1 = 1.6936645$ pF and $C_2 = 1.5903912$ pF giving a $\Delta C = 103.3$ fF. The RCB readout at resonance was 11.76 dBm as shown in Figure 4.

**Table 4. Planar transformer coil measurements at 100KHz**

| Core# | Gap type | Gap (μm) | Winding | $L$ (mH) | $Q$ | $R_S^w$, ESR (Ω) | DCR (Ω) | $L_{leak}$ (μH) | k |
|---|---|---|---|---|---|---|---|---|---|
| C1 | half gap | 65 | Prim(upr) | 4.23 | 127 | 20.5 | 2.61 | 206.20 | 0.98 |
| | | | Prim(lwr) | 4.22 | 132 | 20.1 | 2.62 | 198.60 | 0.98 |
| | | | Secondary | 4.08 | 69 | 37.0 | 2.77 | 87.80 | 0.99 |
| C2 | full gap | 130 | Prim(upr) | 2.55 | 99 | 16.08 | 2.61 | 264.90 | 0.95 |
| | | | Prim(lwr) | 2.56 | 97 | 16.30 | 2.63 | 263.20 | 0.95 |
| | | | Secondary | 2.30 | 79 | 18.00 | 2.60 | 115.50 | 0.97 |
| C3 | full gap | 130 | Prim(upr) | 2.65 | 84 | 19.90 | 2.87 | 262.70 | 0.95 |
| | | | Prim(lwr) | 2.64 | 84 | 19.65 | 2.77 | 265.30 | 0.95 |
| | | | Secondary | 2.39 | 69 | 21.73 | 2.90 | 116.00 | 0.98 |
| C4 | half gap | 65 | Prim(upr) | 4.76 | 111 | 26.80 | 2.73 | 198.27 | 0.98 |
| | | | Prim(lwr) | 4.76 | 114 | 26.30 | 2.74 | 194.29 | 0.98 |
| | | | Secondary | 4.55 | 103 | 28.20 | 2.79 | 85.47 | 0.99 |
| C5 | half gap | 65 | Prim(upr) | 4.32 | 131 | 21.10 | 2.75 | 260.33 | 0.97 |
| | | | Prim(lwr) | 4.29 | 132 | 20.70 | 2.78 | 188.20 | 0.98 |
| | | | Secondary | 4.09 | 112 | 22.50 | 2.95 | 96.82 | 0.99 |
| C6 | ungapped | 0 | Prim(upr) | 40.30 | 20.8 | 1217 | 2.75 | 196.20 | 1.00 |
| | | | Prim(lwr) | 40.62 | 21.2 | 1203 | 2.74 | 203.50 | 1.00 |
| | | | Secondary | 39.10 | 21.7 | 1132 | 2.81 | 86.92 | 1.00 |
| C6 | after thermal cycling[‡] | 0 | Prim(upr) | 46.53 | 19.4 | 1507 | 2.72 | 194.5 | 1.00 |
| | | | Prim(lwr) | 46.69 | 19.7 | 1489 | 2.70 | 201.8 | 1.00 |
| | | | Secondary | 44.69 | 20.4 | 1376 | 2.75 | 87.10 | 1.00 |
| **Averages** | | | | | | | | | |
| ⟨C2, C3⟩ | full gap | 130 | Primary | 2.60 | 91 | 17.98 | 2.72 | 264.03 | 0.95 |
| | | | Secondary | 2.35 | 74 | 19.87 | 2.75 | 115.75 | 0.98 |
| ⟨C1, C4, C5⟩ | half gap | 65 | Primary | 4.43 | 125 | 22.58 | 2.70 | 207.65 | 0.98 |
| | | | Secondary | 4.24 | 95 | 29.23 | 2.84 | 90.03 | 0.99 |
| ⟨C6⟩ | ungapped | 0 | Primary | 43.54 | 20.3 | 1354 | 2.73 | 199.00 | 1.00 |
| | | | Secondary | 41.90 | 21.1 | 1254 | 2.78 | 87.01 | 1.00 |

This power level is equivalent to 0.864 V$_{RMS}$ in a 50 Ω system. Using $f_{conv}^G = 86.4$ μV/fF as the conversion factor at 0.1 V excitation yields a differential capacitance of 100 fF. The difference between the RCB and the bridge measurements is ~ 1.5%. We note that the AH2500A measurement frequency is 1 kHz while the resonant bridge operates at 117.75 kHz which could explain the difference between the two measurements.

---
[‡] Thermal cycling to 120 K and 350 K

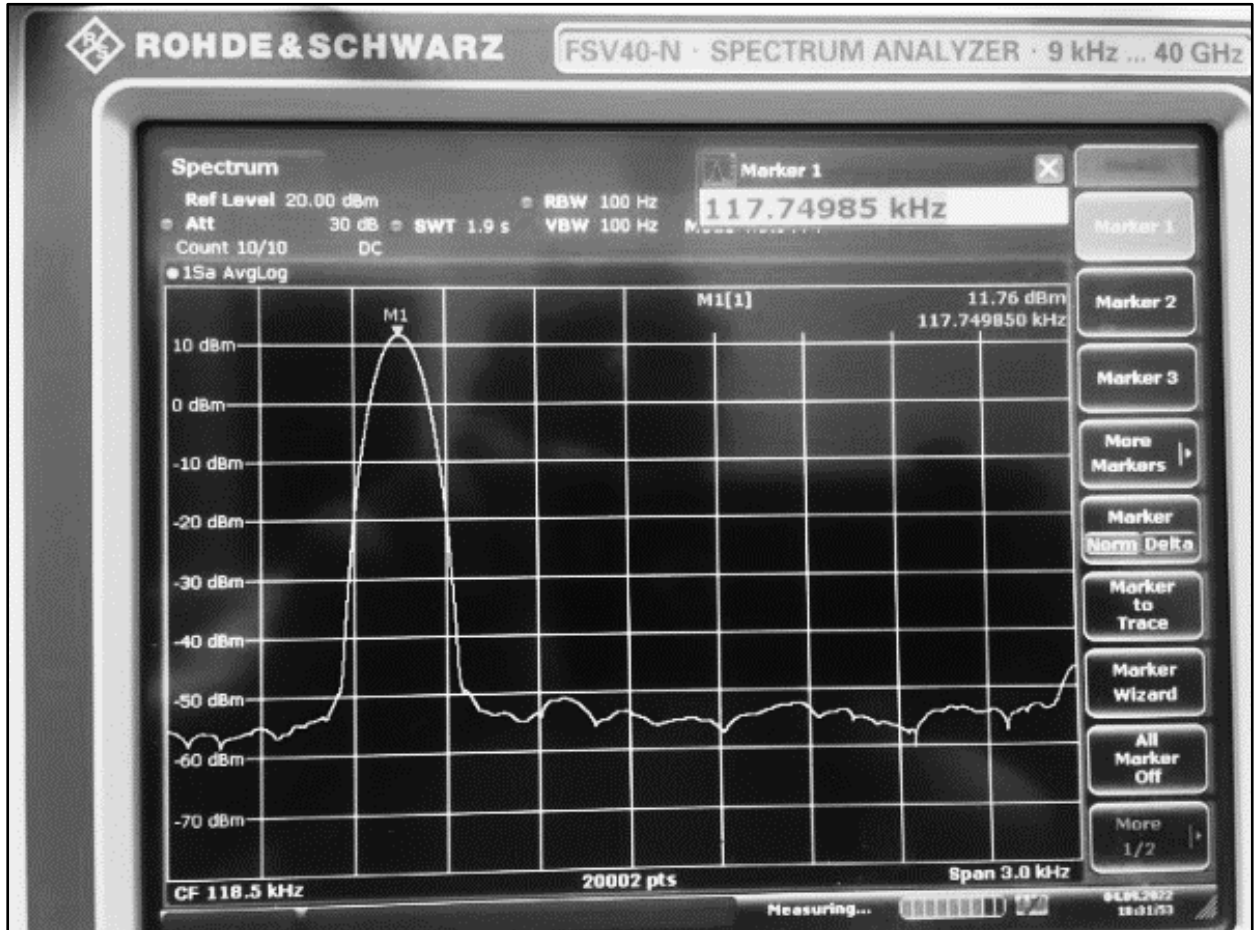

**Figure 4. The RCB output for a frequency scan across the bridge resonant frequency with a 100mV$_{rms}$ excitation input and ~100 fF differential capacitance. Scan: 117 kHz to 120 kHz (0.3 kHz/division)**

### III.C. NOISE FLOOR MEASUREMENTS

The excitation input to the bridge was terminated to ground with a 50 Ω load. A Rohde & Schwarz (R&S) model FSV40-N[35] spectrum analyzer was connected to the output of the RCB for noise floor measurements. The noise floor at 117.75 kHz was measured to be -60.56 dBm at a resolution bandwidth (RBW) of 100Hz with a 100 average count as shown in Figure 5.

Using the conversion factor $f_{conv}^G = 86.4\ \mu V/fF$ (0.1 V excitation), dividing by the square root of RBW, and accounting for the √2 demodulation process noise penalty, $f_{dem} = \sqrt{2}$, yields a sensor sensitivity of about 0.33 aF/√Hz (equation 21 below). This is 50% above the calculated value $S_{min}^{1/2} \cong 0.23$ aF/√Hz. (equation 20) and is due to the use of a core with a gap of $s = 65\mu m$. The RCB noise in aF/√Hz is displayed on the secondary vertical scale on the right of Figure 5. All noise values in the tables are given in both dBm, as measured, and in aF/√Hz, while the figures starting from Figure 6 show only the aF/√Hz scale. For the ungapped core, $S_{min}^{1/2} = 0.22$ aF/√Hz matching the calculated value. Note that, while the $Q$ of the transformer is >100, components of the RCB including the FET amplifiers (with a bandwidth of ≈8 MHz), the feedback resistors, and others introduce losses that drop the overall $Q$ of the circuit by a factor of more than three. The excitation frequency corresponds to a noise minimum in the output, a compelling demonstration of the benefits of resonant operation.

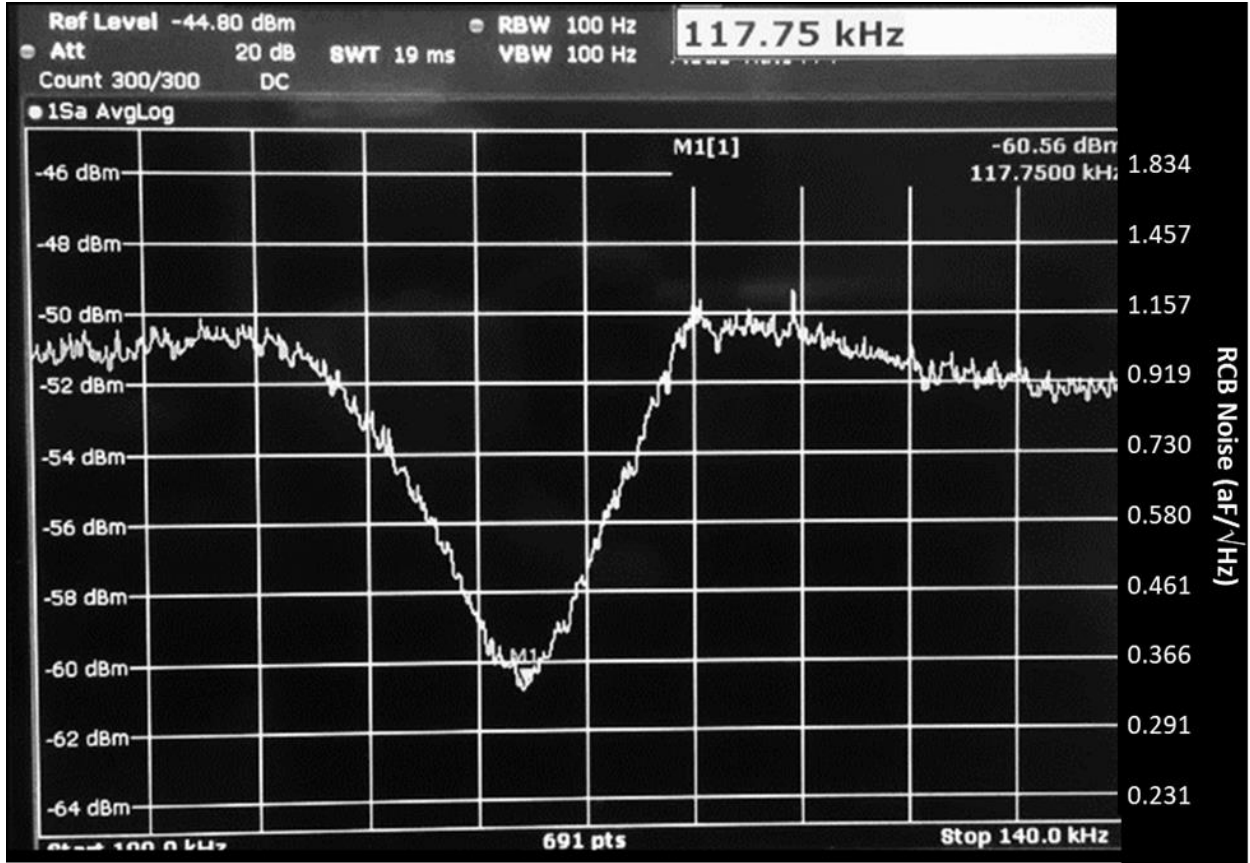

**Figure 5.** Noise floor measurements of a preliminary version of the RCB. The marker (M1) is located at the bridge resonance frequency, the noise minimum of the resonant system. Scan: 100 kHz to 140 kHz (4 kHz/division).

### III.D LONG TERM NOISE FLOOR AND STABILITY

The long-term noise at the bridge resonant frequency was measured using the R&S spectrum analyzer, averaging for 40 s over 400 spectra for 28 hours for cores #1 and #2 (65 μm and 130 μm) and for 40 hours for core #6, with the noise recorded after a continuous quiet measurement period and its resulting stability performance shown in Figure 6. The corresponding frequency domain ranges are 10 μHz to 12.5 mHz and 7 μHz to 12.5 mHz. The noise floor averages, $\langle S_c^{1/2} \rangle$, and their drift rates, $d(\langle S_c^{1/2} \rangle)/dt$, obtained from linear fits to the data in Figure 6 are given in Table 5. The corresponding capacitance measurement sensitivity is given by the conversion from noise power in dBm, $P(\text{dBm})$, to RCB noise $S_c^{1/2}$ (aF/√Hz), by:

$$S_c^{1/2}(\text{aF}/\sqrt{\text{Hz}}) = \frac{\sqrt{\frac{10^{P(dBm)/10}}{1000} R_{50}(\Omega)} f_{dem}}{G_{diff} G_{ext} \sqrt{RBW(\text{Hz})} f_{conv}(\text{V/aF})} = 366 \times 10^{P(dBm)/20} \quad (21)$$

where $R_{50} = 50\Omega$ is the system impedance and the noise voltage is $V_N(V) = \sqrt{\frac{10^{P(dBm)/10}}{1000} R_{50}(\Omega)}$.

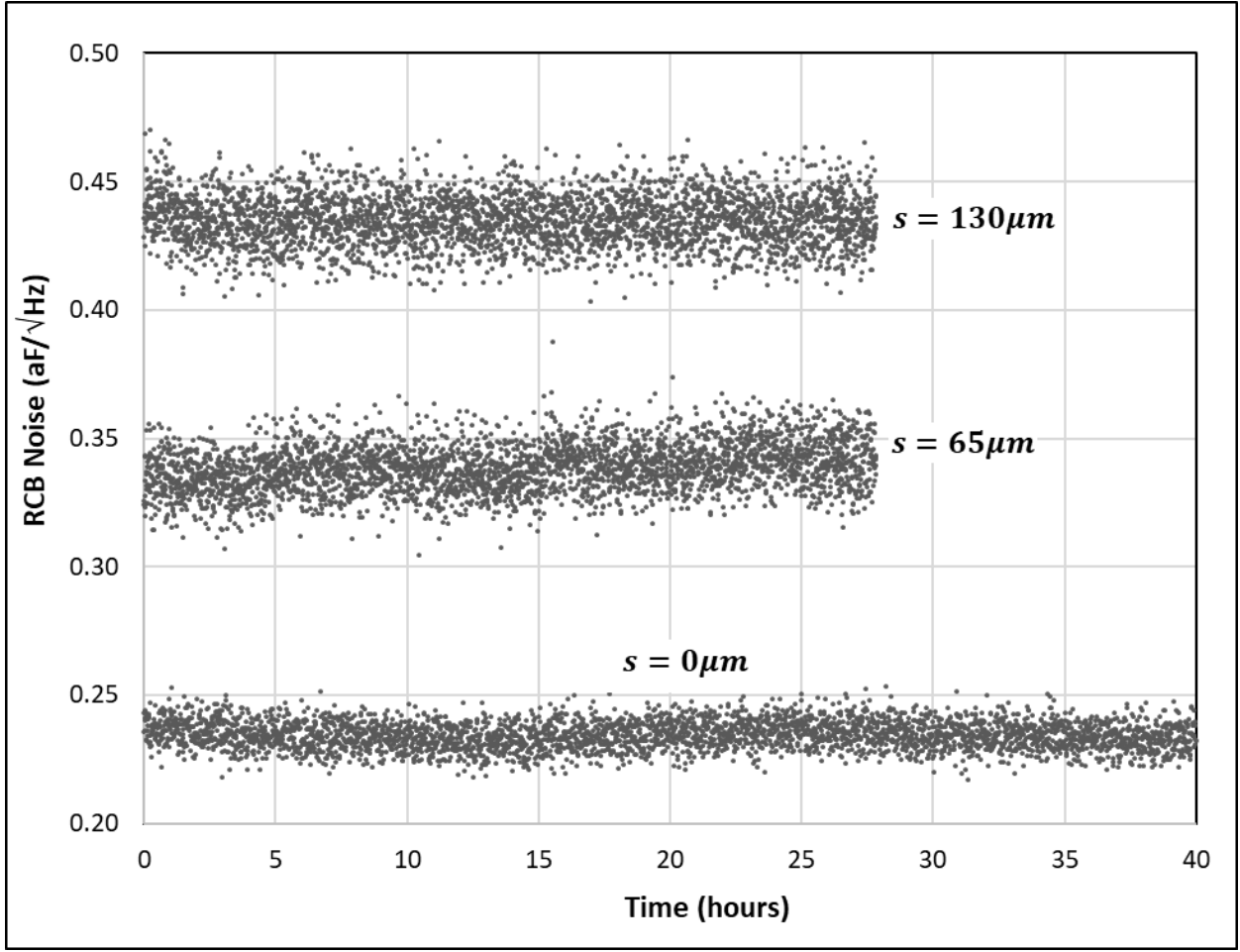

**Figure 6. Long-term noise floor measurements of the RCB for CORE#2 - 130 μm gap (upper data set), CORE#1 - 65 μm gap (middle data set), and CORE#6 – ungapped (lower data set) with a 400-sweep (40 s) average over 28 hours to 40 hours continuous measurement periods.**

**Table 5. Long term stability of cores with core gaps of 130 μm, 65 μm and 0 μm at room temperature, 293 K.**

| Core# | Gap type | Gap (μm) | Duration (hours) | $\langle S_c^{1/2} \rangle$ (dBm) | $\langle S_c^{1/2} \rangle$ (aF/√Hz) | $d(\langle S_c^{1/2} \rangle)/dt$ ((aF/√Hz)/h) |
|---|---|---|---|---|---|---|
| 2 | full gap | 130 | 28 | -58.50 ± 0.20 | 0.435 ± 0.010 | -2×10⁻⁵ |
| 1 | half gap | 65 | 28 | -60.68 ± 0.24 | 0.338 ± 0.009 | 2×10⁻⁴ |
| 6 | Ungapped | 0 | 40 | -63.87 ± 0.20 | 0.234 ± 0.005 | -10⁻⁵ |

Figure 7 shows the data of Figure 6 in the frequency domain, between $10^{-5}$ Hz and $12.5\times10^{-3}$ Hz with a five points data running averaging.

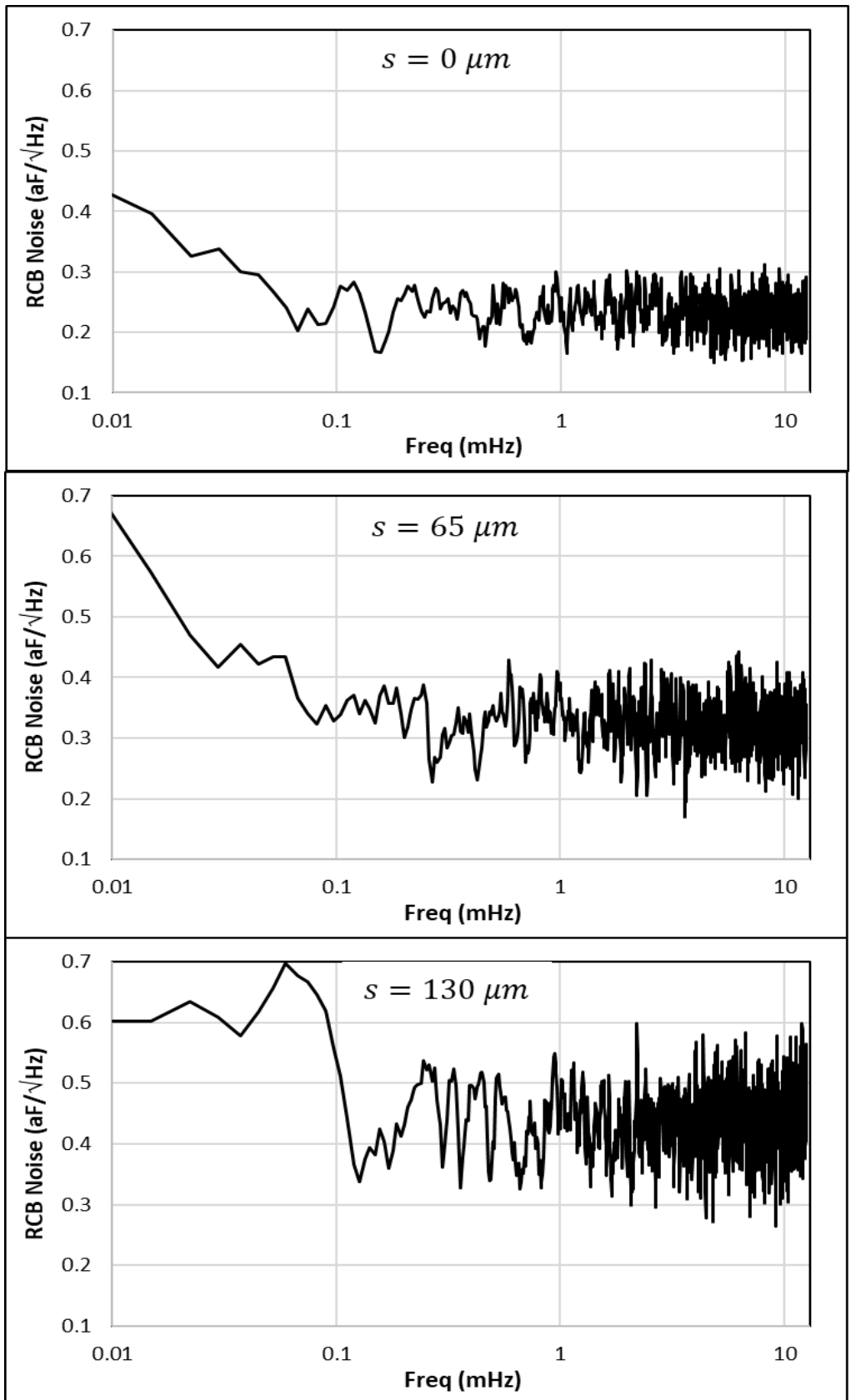

**Figure 7. Long term RCB noise floor in the frequency domain. Data from Figure 6.**

**IV RCB TEMPERATURE DEPENDENT NOISE PERFORMANCE**

As discussed earlier in the paper, the noise floor of the instrument should be ideally determined by the thermal noise of the transformer and should scale as the square root of the temperature. This assumes of course that there are no significant changes in the physical and electrical properties of the components. We expect the bridge $Q$ to improve somewhat at lower temperatures, due to the reduced ohmic losses in the windings that should exceed the increased power losses in the ferrite cores[36,37]. Core losses at low flux density increase between 293 K and 120 K by a factor of order ten while the initial permeability $\mu_i$ decreases by a factor of about five; see for example[36] the data for N87 at 10 kHz, a material similar to N41.

**IV.A CRYOGENIC MEASUREMENTS**

The measurement sensitivity scales as the square root of $T/[f_0^3 L(T)Q]$ as per equation 3. In order to improve the sensitivity of the bridge, the transformers were cooled to 120 K by immersing them in the vapor above liquid Nitrogen (LN2) contained in a dewar flask, thus avoiding the mechanical vibration issues caused by the boiling LN2. The transformer and the NP0 tuning capacitors were installed on a circuit board that was inserted into the dewar containing about one liter of LN2. The transimpedance amplifiers and subsequent amplifier stages were on a separate circuit board outside the dewar in ambient air. The transformer secondary winding was connected to the amplifiers using a twinax cable with the shield driven at the electronic ground of the amplifier system to minimize noise pickup in the laboratory.

Figure 8 shows the noise floor of the instrument with Core #1 at room temperature and at 120 K. Note that for this case the system was retuned when the core was at 120 K, to match the room temperature resonant frequency of 118.195 kHz. The noise reduction is ~4 dB which corresponds to a ~37.5% improvement in the capacitance measurement sensitivity. Thermal noise scales as the square root of the temperature, resulting in an expected ~55% improvement in sensitivity between 293 K and 120 K compared to the measured ~38%. This core suffered cracking damage during cooling, resulting in an anomalous behavior as evidenced by a significant decrease in $Q$ at 120 K.

Extrapolating the thermal noise and lower ohmic losses to 3 K would yield a sensitivity of 0.03 aF/√Hz, which might however require a chemical redesign of the ferrite core that presently has significantly increased losses at low temperature[36]. This is an encouraging prediction and in principle we could measure an energy change of 0.2 eV.

Measurements with Core#2 yielded the results shown in Figure 9 for 293 K, 235 K, 142 K and 120 K. Stabilization times were one hour for 293 K and 120 K, ensuring thermal equilibrium of the system, and resulting in a noise floor improvement from 0.464 aF/√Hz to 0.353 aF/√Hz, an increase in $Q$ from about 24 to 30 and a shift in resonant frequency from about 120 kHz to 132 kHz. For a real signal on the bridge, the two transformer windings will undergo the same change in inductance and the differential signal, to first order, will stay unchanged. However, assuming the NP0 ceramic capacitors have a relatively low temperature coefficient, the resonant frequency will increase at lower temperatures, given that the permeability (and in turn the inductance) decreases as we lower the temperature. From the data in Figure 9 we estimate the temperature coefficient for the resonant frequency to be about 70 Hz/K. The measurements at 235 K and 142 K were performed during active cooldown, and, while demonstrating the trends for the noise floors, the quality factors, and the resonant frequencies, they do not reliably represent the equilibrium parameters of the RCB at these temperatures.

Measurements with cores #3, #4, and #5 are shown in Figure 10. Figure 11 shows the results obtained with Core#6. In addition, Core #6 was cycled to 350 K after it being cooled to 120 K.

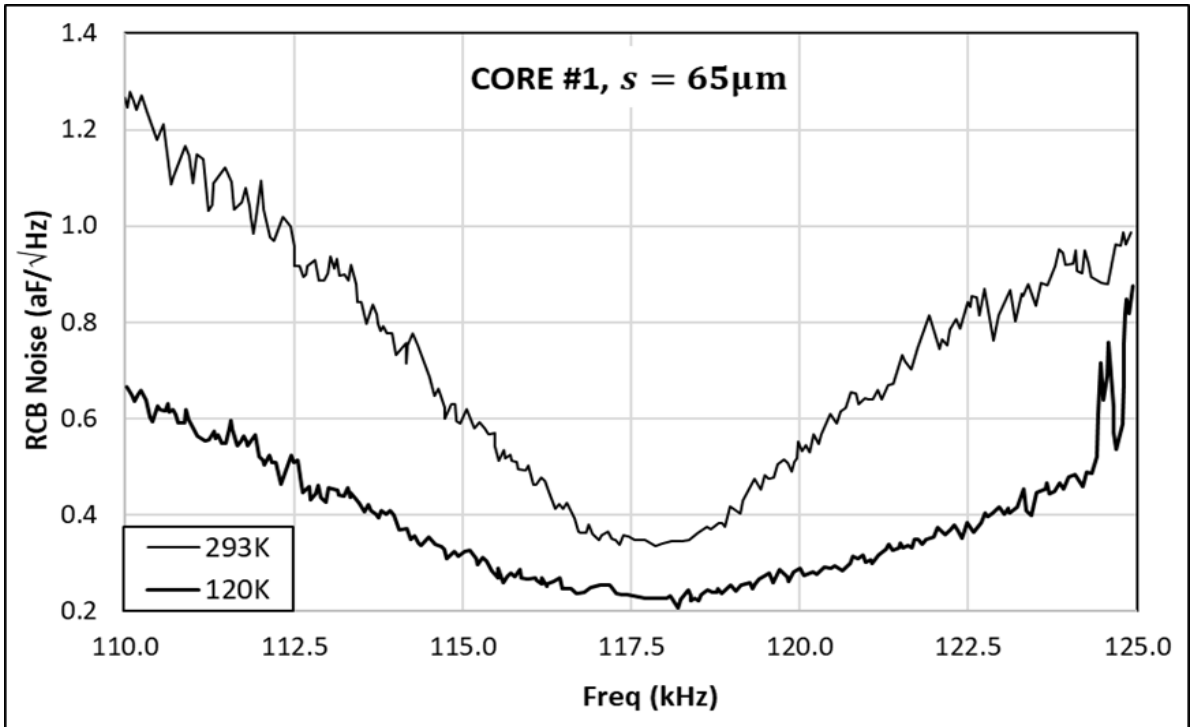

**Figure 8 Noise floor measurements for CORE#1 at room temperature (upper trace) and 120 K (lower trace). with minima corresponding to a detection sensitivity of 0.303 aF/√Hz and 0.223 aF/√Hz, respectively.**

Intermediary noise performance spectra were taken during cooldown, at approximately 25.5 K intervals, though these were measured during cooldown, and thus not at stabilized temperatures.

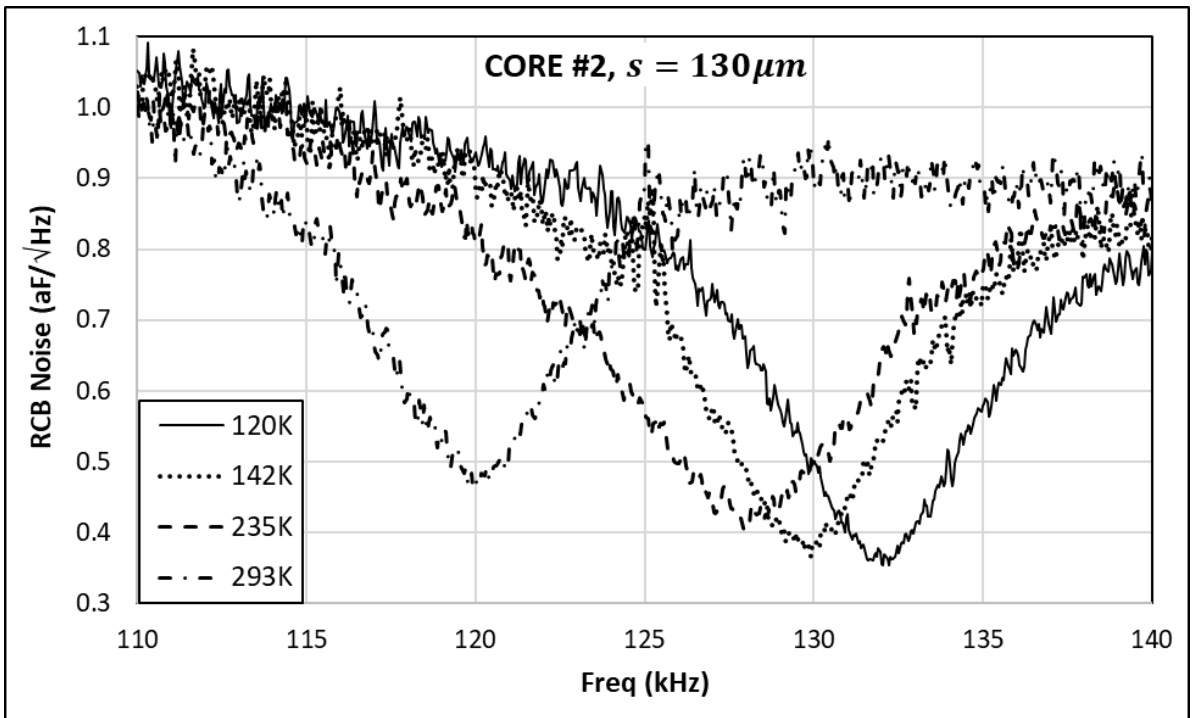

**Figure 9. Noise floor measurements for CORE#2 nominally at 293 K, 235 K, 142 K, and 120 K. The minima correspond to detection sensitivities of 0.464 aF/√Hz, 0.402 aF/√Hz, 0.365 aF/√Hz, and 0.353 aF/√Hz.**

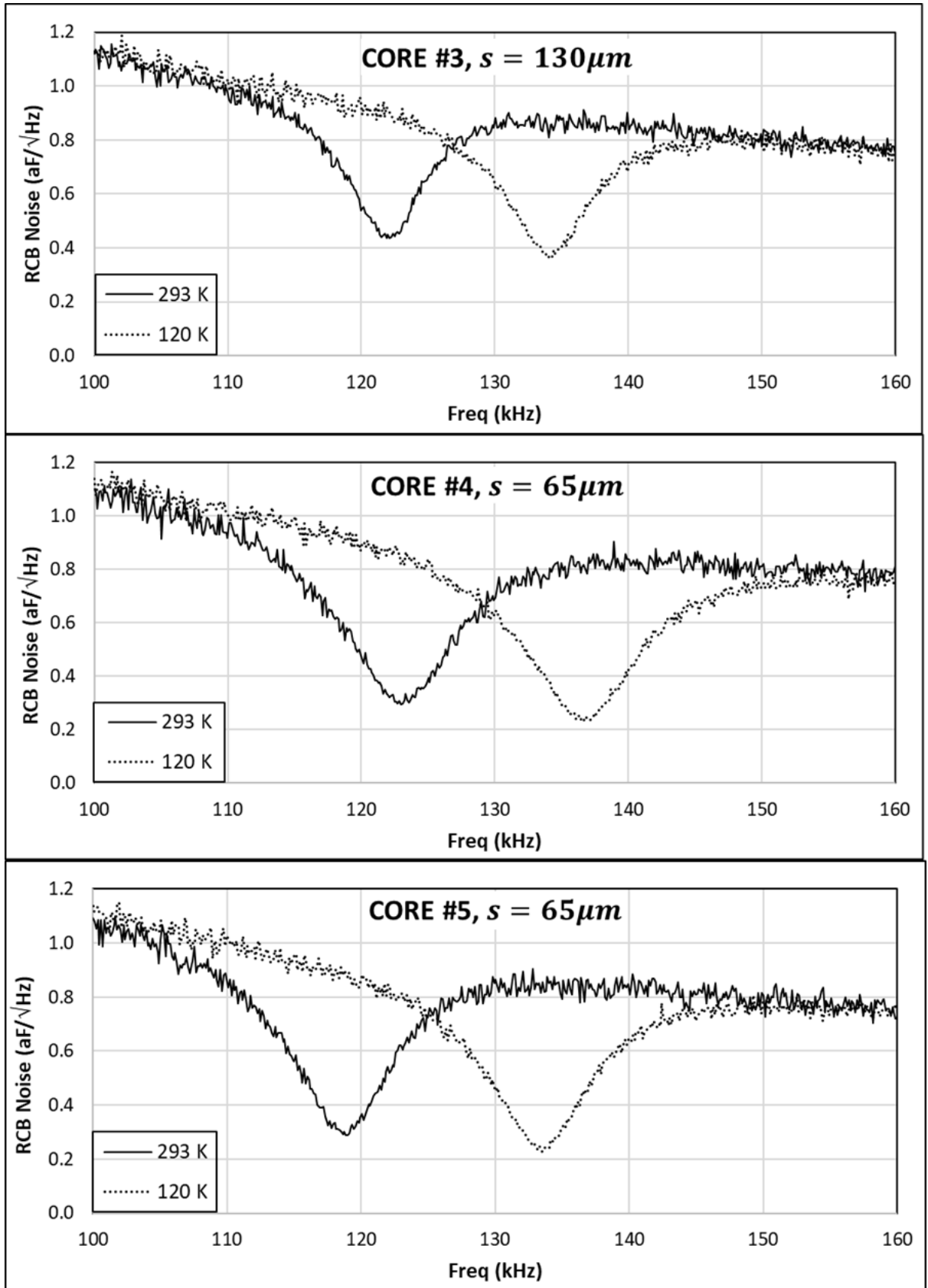

**Figure 10. Noise floor measurements for CORE#3, CORE#4, and CORE#5 at 293 K and 120 K.**

During the cooldown the capacitance noise decreased monotonically, as expected, with the measured values recorded in Table 8.

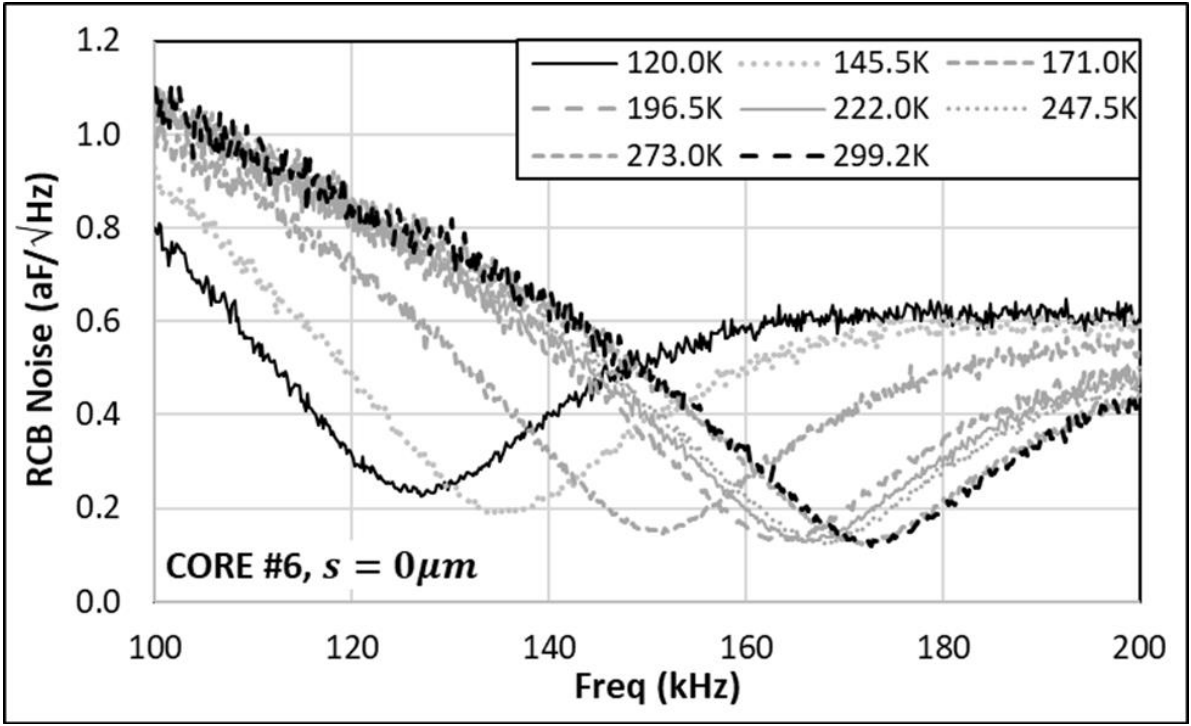

**Figure 11. Noise floor measurements for CORE#6. Solid and dashed black traces are measured at 293 K and 120 K. Gray traces are taken at intermediate temperatures at approximately 25.5 K intervals during cooldown.**

Table 6 gives a summary of the RCB noise minimum, $S^{1/2}$, at 293 K and 120 K for all six cores in units of dBm, as measured, and in aF/√Hz, the corresponding capacitance bridge noise. Also listed are $f_0$, the RCB resonance frequency at which $S^{1/2}$ is measured, and $f_{3db}$, the frequency where the noise value is 3db higher than $S^{1/2}$, that is used in the calculation of the quality factor $Q$, whose values are also included. We mention once more that the anomalous results for Core#1 at low temperature are due to ferrite core cracking and retuning the bridge so that $f_0^{120K} = f_0^{293K}$.

**Table 6. RCB performance at 293 K and 120 K for the six cores.**

|  | C1 (65μm) | | C2 (130μm) | | C3 (130μm) | | C4 (65μm) | | C5 (65μm) | | C6 (0μm) | |
| --- | --- | --- | --- | --- | --- | --- | --- | --- | --- | --- | --- | --- |
|  | 293K | 120K | 293K | 120K | 293K | 120K | 293K | 120K | 293K | 120K | 293K | 120K |
| $S^{1/2}$ (dBm) | -61.08 | -65.07 | -57.9 | -60.32 | -58.47 | -60.03 | -61.93 | -63.90 | -62.07 | -64.14 | -64.24 | -69.82 |
| $S^{1/2}$ (aF/√Hz) | 0.323 | 0.204 | 0.464 | 0.353 | 0.444 | 0.371 | 0.298 | 0.237 | 0.293 | 0.231 | 0.225 | 0.118 |
| $f_0$ (kHz) | 118.2 | 118.2 | 119.9 | 132.0 | 121.9 | 134.0 | 123.0 | 136.4 | 118.9 | 133.5 | 127.1 | 172.7 |
| $f_{3db}$ (kHz) | 116.2 | 115.2 | 117.2 | 129.8 | 119.5 | 131.9 | 120.8 | 134.6 | 116.7 | 131.9 | 119.4 | 167.9 |
| Q | 29.6 | 19.7 | 22.2 | 30.0 | 25.3 | 31.0 | 28.4 | 37.8 | 27.5 | 39.7 | 8.3 | 18.0 |

Table 7 gives the scaling of the principal performance parameters of the bridge, for all six cores, as a function of temperature between 293 K and 120 K (in dBm as measured and in aF/√Hz), with the first row showing $(df_0/f_0)/dT$, the variation of $f_0$ with temperature. The average values for full gap, Core#2 and #3, for half gap, Core#4 and #5, and for no gap, Core#6 are:

$$\langle (df_0/f_0)/dT \rangle_{120} = -.55 K^{-1}; \quad \langle (df_0/f_0)/dT \rangle_{65} = -.66 K^{-1}; \quad \langle (df_0/f_0)/dT \rangle_0 = -1.76 K^{-1} \quad (22)$$

The second and third rows of Table 7 give the difference in the minimum noise in dBm and its ratio when converted to aF/√Hz. In rows four and five of Table 7 we calculate the ratios of the quality factors $Q$ and of the frequencies of minimum noise $f_0$.

Table 7. Scaling of principal RCB parameters for cores #2, #3, #4, #5 and #6 between 293 K and 120 K.

|  | C1 | C2 | C3 | C4 | C5 | C6 |
|---|---|---|---|---|---|---|
| $(df_0/f_0)/dT$ (K$^{-1}$) | N/A | -.555 | -.546 | -.597 | -.669 | -1.76 |
| $S_{120K}^{1/2}(\text{dBm}) - S_{293K}^{1/2}(\text{dBm})$ | -4.01 | -2.42 | -1.56 | -1.97 | -2.07 | -5.58 |
| $S_{120K}^{1/2}(\text{aF}/\sqrt{\text{Hz}})/S_{293K}^{1/2}(\text{aF}/\sqrt{\text{Hz}})$ | 0.63 | 0.76 | 0.84 | 0.80 | 0.79 | 0.52 |
| $Q^{120K}/Q^{293K}$ | N/A | 1.35 | 1.22 | 1.33 | 1.45 | 2.17 |
| $f_0^{120K}/f_0^{293K}$ | N/A | 1.11 | 1.10 | 1.11 | 1.12 | 1.36 |

The initial permeability decreases significantly with temperature. Figure 12 displays the dependence of the permeability $\mu_i$ over the temperature range 93 K to 470 K, using the combined experimental data for N87[36] and the TDK supplied values for N41[26]. N41 and N87 material values, including $\mu_i$, overlap in within the manufacturer's specifications tolerances. Effective permeabilities for half and full gapped cores, as calculated from equation 5, are also shown. The ratios of permeability values between 293 K and 120 K decreases with increasing gap value:

$$\mu_i^{0\mu/293}/\mu_i^{0\mu/120} = 5.53, \quad \mu_e^{65\mu/293}/\mu_e^{65\mu/120} = 1.97, \quad \mu_e^{130\mu/293}/\mu_e^{130\mu/120} = 1.54 \quad (23)$$

Note the very large variation in the rate of change of the permeability over the range of temperatures 120 K to 293 K encompassing the cryogenic tests of the RCB:

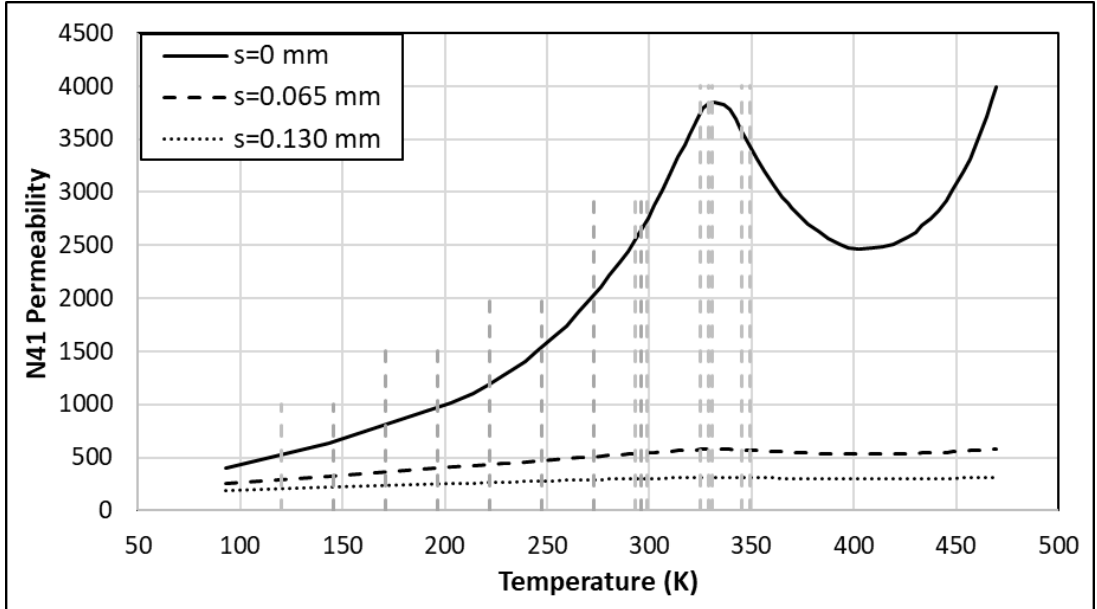

Figure 12. N41/N47 core material permeability as function of temperature. Data for $s = 0$ mm from references 26 and 36. The vertical gray dashed lines mark all the temperatures at which noise spectra were measured.

$$\left(\Delta\mu_e^{0\mu}/\Delta T\right)_{120K}^{293K} = 11.8K^{-1}, \quad \left(\Delta\mu_e^{65\mu}/\Delta T\right)_{120K}^{293K} = 1.5K^{-1}, \quad \left(\Delta\mu_e^{130\mu}/\Delta T\right)_{120K}^{293K} = 0.6K^{-1} \quad (24)$$

While $\mu_i$ for the ungapped core and its temperature derivative over this range vary strongly with temperature, the gapped core equivalent dependence is greatly attenuated. The temperature derivative is practically constant for the gapped cores as seen from their averages and their corresponding standard deviations:

$$\langle (d\mu_e^{65\mu}/dT)_{120K}^{293K}\rangle = 1.53 \pm 0.22 \text{ K}^{-1}, \quad \langle (d\mu_e^{130\mu}/dT)_{120K}^{293K}\rangle = 0.59 \pm 0.13 \text{ K}^{-1} \quad (25)$$

**IV.B RCB NOISE MEASUREMENTS ABOVE ROOM TEMPERATURE**

Figure 13 shows the RCB noise for the ungapped core #6 for six temperatures between 299.2 K and 348.6 K.

For the ungapped Core #6, Figure 14 displays from 299 K to 349 K in the left panel the dependence of the noise floor minima values with a linear fit to temperature, while the right panel shows the frequencies corresponding to these minima and their fit to $1/\sqrt{\mu}$ for the N41 MnZn.

The RCB the resonant frequency $f_0$ of equation 1 is also the frequency of the noise minimum and, as $C_{eq}$ can be considered temperature invariant, we obtain the following approximation:

$$f_0 \widetilde{\propto} 1/\sqrt{L(T)} \widetilde{\propto} 1/\sqrt{\mu(T)} \quad (26)$$

In the right panel of Figure 14 we fit the permeability data from Figure 13 to the data points for the frequencies of the minima of the RCB capacitance noise. The data properly reproduces the maximum in the permeability curve of the MnZn N41 material, thus validating the approximation in equation 26.

Equation 3 gives the RCB noise minima as a function of $T, f_0, L, Q$ and using equation 26 we obtain:

$$S_C^{1/2} \widetilde{\propto} \sqrt{T/[f_0(T)Q(T)]} \widetilde{\propto} \sqrt{T/Q(T)} \quad (27)$$

where, as seen above, while $f_0(T)$ varies with the permeability, its peak-to-valley variation is $\leq 4\%$ allowing the approximation $f_0(T) \cong$ constant, and resulting in the second proportionality factor of equation 27.

$Q$ decreases with temperature, as per equation 18 and as observed from Table 8, and consequently the noise floor increase with temperature faster than $\sqrt{T}$. The data in the left panel of Figure 14 shows an approximate linear fit for $S_C^{1/2}$ with temperature.

Table 8 summarizes all the measurements for the ungapped Core #6 between 120.0 K and 348.6 K. The shaded columns represent measurements during cooldown, at which the temperatures did not necessarily reach equilibrium. As expected, the value of $Q$, measured in within approximately ±5%, decreases with increasing temperature.

**Table 8. Performance values for the ungapped Core #6 at temperatures between 120.0 K and 348.6 K.**

| Temp (K) | 120.0 | 145.5 | 171.0 | 196.5 | 222.0 | 247.5 | 273.0 | 293.2 | 296.2 | 299.2 | 325.1 | 329.1 | 330.7 | 345.3 | 348.6 |
|---|---|---|---|---|---|---|---|---|---|---|---|---|---|---|---|
| $S^{1/2}$ (dBm) | -69.8 | -69.5 | -69.4 | -69.1 | -69.1 | -68.0 | -65.9 | -64.2 | -64.3 | -63.9 | -62.4 | -62.1 | -62.0 | -61.2 | -61.2 |
| $S^{1/2}$ (aF/√Hz) | 0.118 | 0.123 | 0.124 | 0.128 | 0.129 | 0.147 | 0.185 | 0.225 | 0.225 | 0.234 | 0.278 | 0.287 | 0.291 | 0.318 | 0.318 |
| $f_0$ (kHz) | 172.7 | 171.7 | 167.9 | 166.1 | 163.9 | 151.5 | 136.5 | 127.1 | 125.9 | 121.0 | 116.1 | 115.9 | 115.9 | 119.1 | 119.9 |
| $f_{-3db}$ (kHz) | 167.9 | 166.5 | 162.7 | 161.1 | 158.9 | 145.5 | 129.3 | 119.4 | 118.0 | 114.8 | 108.1 | 108.4 | 107.5 | 110.7 | 111.3 |
| Q | 18.0 | 16.5 | 16.1 | 16.6 | 16.4 | 12.6 | 9.5 | 8.3 | 8.4 | 8.2 | 6.6 | 7.1 | 5.9 | 5.6 | 5.5 |

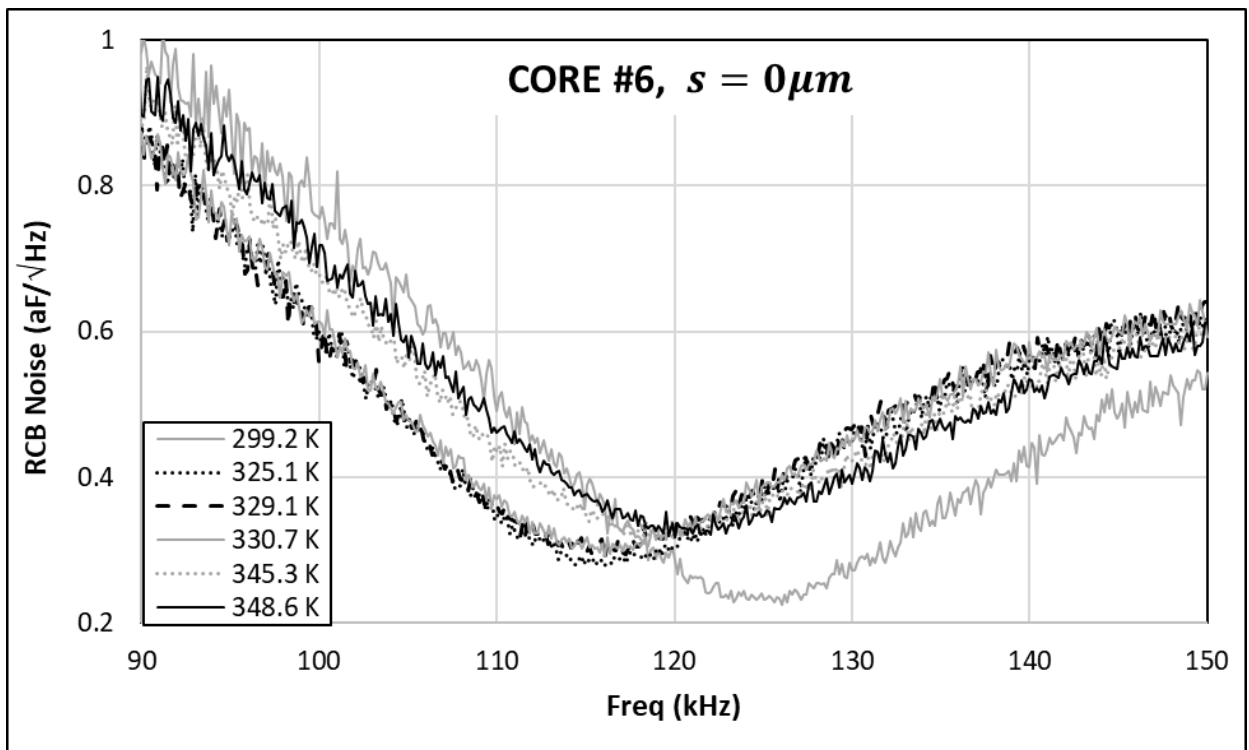

Figure 13. Noise floor measurements for the ungapped CORE #6 between 299.2 K and 348.6K.

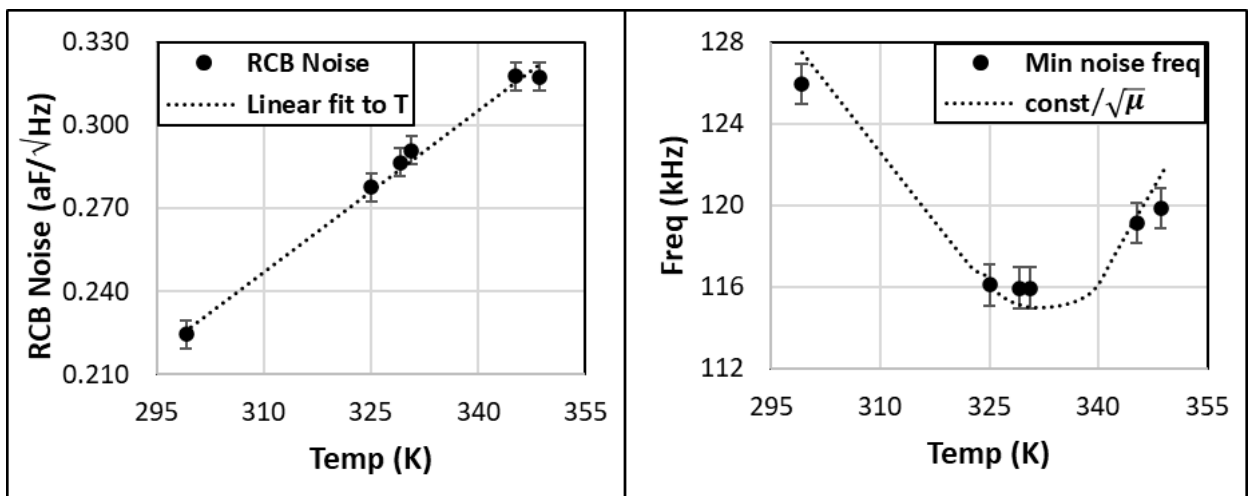

Figure 14. CORE #6 between 299.2 K and 348.6K. *Left*: Minimum RCB noise floor as a function of temperature and linear fit. *Right*: Frequency of the minimum RCB noise floor and a fit to $\text{const}/\sqrt{\mu}$ for MnZn N41.

## V. CONCLUSIONS

We have designed, modelled, and fabricated the front-end of a precision capacitance bridge with a measured long-term average sensitivity, at room temperature for runs of 40 hours for the ungapped core and of 28 hours with one core of each of the two gap types, demonstrated the stability of their long-term sensitivities of $0.234 \pm 0.005$ aF/√Hz, $0.338 \pm 0.009$ aF/√Hz, and $0.435 \pm 0.010$ aF/√Hz for the ungapped, 65 μm gap, and 130 μm gap cores respectively. The differential transformer at the heart of the resonant capacitance bridge was fabricated using planar windings on a FR4 PCB and MnZn N41 ferrite cores with excellent repeatability, as demonstrated with five more cores. Further improvements in the transformer $Q$ and overall capacitance sensitivity can be achieved by using ceramic substrates with lower $\varepsilon_r$ and lower loss for the windings. The best noise performance was obtained with the ungapped core configuration; $0.225 \pm 0.005$ aF/√Hz and $0.118 \pm 0.005$ aF/√Hz at 293 K and 120 K, a 48% noise reduction. Cooling the three 65 μm gap transformers to 120 K improved their bridge sensitivities by about 27% from an average detection threshold of $0.30 \pm 0.01$ aF/√Hz to one of $0.22 \pm 0.01$ aF/√Hz. Two transformers with 130 μm core gaps yielded sensitivity averages of $0.45 \pm 0.01$ aF/√Hz and $0.36 \pm 0.01$ aF/√Hz at 293 K and 120 K, thus showing a 20% noise reduction. The room temperature floor noise for the ungapped core equaled the calculated value of 0.23 aF/√Hz for its RCB configuration. Resonant frequency temperature variation for the 65 μm and the 130 μm cores averaged to about -70 Hz/K and -81 Hz/K respectively, much less than the ungapped case value of -261 Hz/K. We also confirmed the consistent scaling with temperature of the parameters of the RCB for all six cores. Measurements with the ungapped core between 300 K and 350 K further validated the expected dependence of the RCB noise floor value on temperature and of the frequency of the noise floor on the permeability of the core material. The corresponding frequency range for the 40-hour room temperature noise measurement time series of the un-gapped RCB is 7 μHz to 12.5 mHz, while that of the 28 hour range for the 65 μm and 130 μm gaps is 10 μHz to 12.5 mHz. Our measurements confirm the analytical expectation that there is a clear trade-off between low noise performance and the temperature coefficient of the noise minimum versus frequency for resonant transformer bridge circuits. The bridge easily meets the demanding requirements of a position sensor for a drag-free gravitational reference sensor[9].

## ACKNOWLEDGEMENTS

SN&N Electronics acknowledges funding from the Changchun Institute of Optics, Fine Mechanics and Physics (CIOMP) for the feasibility study and subsequent fabrication and test of an early model of the precision capacitance bridge at room temperature. The authors thank Stanford University for the use of test equipment.